\documentclass{revtex4}

\usepackage{amsmath, amssymb, graphicx, comment, amsthm}
\usepackage{bbold, bm}
\usepackage{color}

\newcommand{\req}{Eq.\ref}
\newcommand{\none}{\nonumber\\}
\newcommand{\be}{\begin{eqnarray}}
\newcommand{\ee}{\end{eqnarray}}
\usepackage{hyperref}
\usepackage{epsfig} 
\usepackage{graphicx}
\usepackage{amsfonts}
\usepackage{amssymb}
\usepackage{amsbsy}
\usepackage{amsmath}
\usepackage{latexsym}
\usepackage{dsfont}
\usepackage{comment}

\def\none{\nonumber\\}

\def\bea {\begin{eqnarray}}
\def\eea {\end{eqnarray}}
\def\F {$|\! +\! i +\! i +\! i\rangle \ $}
\def\I {$|\! +++\rangle \ $}
\def\cb{{}}

\def\eightvector#1#2#3#4#5#6#7#8{ \left(
\begin{tabular}{c}
$#1$\\
$#2$\\
$#3$\\
$#4$\\
$#5$\\
$#6$\\
$#7$\\
$#8$\\
\end{tabular}
\right)}
\def\bra#1#2#3{| #1#2#3\rangle}

\begin{document}

\title{ Escape from the Quantum Pigeon Conundrum }
\author{Gabor Kunstatter}
\email{g.kunstatter@uwinnipeg.ca}
\affiliation{University of Winnipeg\\
Physics Department\\
Winnipeg, MB Canada \ R3B 2E9}
\author{Jonathan Ziprick}
\email{jziprick@rrc.ca}
\author{Victoria McNab}
\author{Alexander Rennie}
\author{Connor Speidel}
\author{Jovin Toews}
\affiliation{Red River College\\
Applied Computer Education Department\\
Winnipeg, MB Canada \ R3B 1K9\\}
\date{\today}

\begin{abstract}
It has recently been argued in Aharonov et. al. (2016) that quantum mechanics violates the Pigeon Counting Principle (PCP) {which states that} if one distributes three pigeons among two boxes there must be at least two pigeons in one of the boxes. {\cb However, this conclusion {cannot justified by rigorous theoretical arguments}. The issue is further complicated by experimental confirmation of the transition amplitudes predicted in this paper that {nevertheless} do not support the conclusion of PCP violation. {Here we prove via a set of operator identities} that the PCP is {not violated} within quantum mechanics, regardless of interpretation.}
\end{abstract}

\maketitle

\tableofcontents
\clearpage

\section{Introduction}
\label{sec:Introduction}

There are many examples of experimentally verified quantum predictions that appear to threaten our intuition concerning the nature of reality. Some of the most startling are based on Bell's inequalities \cite{Bell1964}, which show that in the case of certain entangled but spatially separate systems, the statistical predictions of quantum mechanics forbid one from assigning specific values (referred to as \textit{hidden variables}) to quantum observables even when these quantities appear to satisfy the  natural criteria for elements of reality posited by Einstein, Podolsky and Rosen (EPR) \cite{EPR}. More to the point, hidden variables are forbidden in these cases unless one violates locality or allows that the sum of two positive numbers can be negative \footnote{See \cite{Mermin1981}, \cite{Kunstatter1984} for accessible reviews.}.

Recently, Aharonov et. al. \cite{ACPSST} argued that a seemingly fundamental principle of Nature known as the pigeonhole counting principle (PCP) is violated in quantum mechanics. The classical PCP is the intuitive statement that if one puts three pigeons in two boxes, then there must be at least two pigeons in one of the boxes. Aharonov et. al. consider the quantum mechanics of three qubits which play the role of pigeons, and argue that for specific pre-selected initial and post-selected final pigeon states (\I and \F respectively) there is a violation of the PCP, i.e. no two qubits are in the same `hole' ($|0\rangle$ or $|1\rangle$).  The authors conclude that this effect sheds ``new light on the very notions of separability and correlations in quantum mechanics and on the nature of interactions''. We refer to this puzzle as the quantum pigeonhole conundrum (QPC).

The arguments in \cite{ACPSST} can be summarized in three mains steps \footnote{Svensson has published a letter \cite{Svensson} arguing against these conclusions and that a rebuttal by Aharonov et. al. \cite{ACPSST2} was published in response. We present different arguments here.}:
\begin{enumerate}
\item The authors consider the projection operator $\Pi_{ab}$ onto the sub-space of three pigeon states for which pigeons $a$ and $b$ are in the same box, and show that
\label{APCSST1}
\bea
 \langle + i +\! i +\! i | \Pi_{ab} |+ + + \rangle =0; \qquad \hbox{for all} \quad \{a,b\} \in \{ 0, 1, 2 \}.
 \label{eq:PiabTransition}
 \eea
 This in turn implies that the following transition amplitude also vanishes:
 \bea
 \left| \langle + i +\! i +\! i | \Pi | +++\rangle \right|^2 =0,
 \label{eq:PiTransition}
 \eea
 where 
 \bea
\Pi := \Pi_{01} + \Pi_{12} +\Pi_{20}.
 \eea
 
 \item Aharonov et. al. then consider a thought experiment where an ensemble of quantum 
 \label{APCSST2} 
 three-particle states, with each element in the pre-selected initial state \I at $t_i$ and post-selected final state \F at $t_f$. They argue that (\ref{eq:PiabTransition}) implies post-selection eliminates all elements of the ensemble for which $\Pi |+ + + \rangle  \neq 0$, i.e. post-selection removes all pre-selected elements with two qubits in the same state. Thus they conclude that, for these particular pre- and post-selected elements of the ensemble, 
``no two particles are in the same box'' during the transition {\it even when the $\Pi_{ab}$ operations are not applied} \footnote{The phrase in italics is our addition.}. 
 
 \item The authors propose that (\ref{eq:PiTransition}) and hence the QPC can be verified experimentally using an electron Mach-Zehnder interferometer to split three weakly 
\label{APCSST3} interacting electrons (quantum pigeons) between two channels. The experiment is based on the weak coupling expansion of an interaction Hamiltonian $H_{int}= (\epsilon t ) \Pi$, with the interaction assumed to be of sufficiently short range such that measurable interaction occurs only when electrons pass through the same channel.  Equation (\ref{eq:PiTransition}) is verified by the absence to first order in $\epsilon t$ of interactions between the electrons in elements of the ensemble that end up in the post-selected state. The vanishing of the transition amplitude (\ref{eq:PiTransition}) has been confirmed in \cite{AHM} using three nuclear qubits as pigeons and in \cite{Chen} using three single photons as pigeons; these authors state that their results confirm no two pigeons are in the same box during the interaction to first order in the coupling. 
\end{enumerate}

{\cb Here we argue that the theoretical arguments summarized in the first two points above do not support the conclusion that the PCP is violated in quantum mechanics. The issue is further confused by published experimental verification of the theory which does not support the conclusion of PCP violation. Our arguments do not simply constitute a different interpretation; we use operator identities to prove mathematically that the PCP is \textit{not violated} in quantum mechanics, regardless of the interpretation, because these identities prevent the relevant observables from being assigned PCP-violating values. These operator identities imply a quantum counting principle that provides a quantum information version of the PCP.

We begin in the next section with a detailed theoretical analysis of the QPC and the electron Mach-Zehnder interferometer experiment, {providing new results that present exact solutions for the time-dependent interference featuring in the transition amplitudes.} In section 3 we present new experimental results for these transition amplitudes obtained using a five-qubit quantum processor through the IBM Quantum Experience cloud service \cite{IBMQ} to {illustrate} the process that is carried out in experiments \cite{AHM, Chen}. {Section 4 contains the main result of our paper by proving that the  apparent violation of the PCP is a result of untenable hidden variable assumptions analogous to those made by EPR \cite{EPR}.} Specifically, we show that the physical Hilbert space is spanned by states each having at least two pigeons in the same box --- {this fact leads to an operator identity that, for any quantum state, forbids assigning a value of zero to all three projection operators $\Pi_{ab}$, as would be required for a violation of the PCP.} Thus, we conclude that it is incorrect to claim that ``there are no two particles in the same box'' during the transition between pre- and post-selected states, and this conclusion is independent of the interpretation one puts on the quantum wavefunction.}

\section{States and operators}
Here we establish notation and define operators that are relevant to the discussion.
We consider the three pigeons to be three electrons, each of which can be found in one of two states (pigeonholes): spin up or down along the $z$-axis. In order to make it easier to connect with our experiment using qubits on a quantum computer processor, we denote the spin-up and spin-down states by $|0\rangle$ and $|1\rangle$ respectively.

There are different bases used in the experiment, and these are related by
\be
|\pm \rangle = \frac{1}{\sqrt{2}}(|0\rangle \pm |1 \rangle), \qquad \qquad | \! \pm \! i \rangle = \frac{1}{\sqrt{2}} ( |0 \rangle \pm i |1 \rangle ) .
\ee
In the pre-selected state $|\! +++\rangle$ all three electron are in an eigenstate of $\sigma_x$ with eigenvalue +1:
\bea
|\! +++\rangle &=& \frac{1}{\sqrt{2}}\left(|0\rangle+|1\rangle\right) \otimes \frac{1}{\sqrt{2}}\left(|0\rangle+|1\rangle\right) \otimes \frac{1}{\sqrt{2}}\left(|0\rangle+|1\rangle\right)\none
&=& \frac{1}{2\sqrt{2}}\left(
|000\rangle + |001\rangle+|010\rangle + | 011\rangle + |100\rangle +|101\rangle + |110\rangle + |111\rangle
\right) .
\label{eq:InitialState}
\eea
In the post-selected state \F all three electrons are  in an eigenstate of $\sigma_y$ with eigenvalue +1:  
\bea
|\! +\! i +\! i +\! i\rangle &=& \frac{1}{\sqrt{2}}\left(|0\rangle+i|1\rangle\right) \otimes \frac{1}{\sqrt{2}}\left(|0\rangle+i|1\rangle\right) \otimes \frac{1}{\sqrt{2}}\left(|0\rangle+i|1\rangle\right)\none
&=& \frac{1}{2\sqrt{2}}\left(
|000\rangle + i|001\rangle+i|010\rangle - | 011\rangle + i |100\rangle - |101\rangle - |110\rangle - i |111\rangle
\right) .
\label{eq:FinalState}
\eea

Of primary interest are the projection operators onto states in which two pigeons are in the same hole, whether it be $|0\rangle$ \textit{or} $|1\rangle$. It takes the form:
\bea
\Pi_{ab}&=&
\mathbb{1}_c \otimes \left(|0_b 0_a \rangle \langle 0_b  0_a | + |1_b 1_a \rangle \langle 1_b 1_a | \right) 
\eea
where $\{a, b, c \}$ is a permutation of $\{0, 1, 2\}$. These projectors do not specify which hole pigeons $a$ and $b$ are in, only that it is the same for both.

Using the definitions above it is easily verified that
\bea
\langle \! +\! i +\! i +\! i | \Pi_{ab} |\! +++ \rangle =0,
\eea
which is the central equation in \cite{ACPSST} and leads to the vanishing of
\be
\langle +i +i +i|\Pi|+++\rangle=0.
\ee

Exploring deeper, we note that the projection operators $\Pi_{ab}$ commute, and the product of any two is also a projector:
\be
\Pi_{01} \Pi_{02} = \Pi_{01} \Pi_{12} = \Pi_{02} \Pi_{12} = P,
\label{eq:PiProduct}
\ee
where
\be
P := |000\rangle \langle 000 | + | 111 \rangle \langle 111 |
\ee
is the projector onto the subspace where all three electrons follow the same path. Using this we can define the projector
\be
p_{ab} := \Pi_{ab} - P
\label{eq:pPi1}
\ee
onto states where electrons $a$ and $b$ follow the same path but $c$ follows the other path, and
\bea
p:= p_{01}+p_{12}+p_{02}
\eea
which projects onto the subspace of states where exactly one pair of electrons follows the same path. The three operators
$p_{ab}$, $p$ and $P$ are all projectors and so have eigenvalues $0$ and $1$.

Importantly, these operators satisfy the following identities:
\bea
\mathbb{1} &=& p +P,
\label{eq:Identity1}\\
\mathbb{1} &=& \Pi -2P.
\label{eq:Identity2}
\eea
Note that $\Pi$, which is the sum of projection operators, is not itself a projection operator because the sum of all the $\Pi_{ab}$ overcounts the states leaving three copies of the subspace in which all three particles are in the same box. This is why subtracting $2P$ as in \ref{eq:Identity2} leaves the identity operator. We note for future reference that the eigenvalues of $\Pi$ are 1 and 3.

The first equation shows that identity operator can be written as the sum of $p$ (with only two electrons following the same path) and $P$ (with all three electrons following the same path). This means that any state in the Hilbert space of the quantum pigeon hole problem can be written as a linear combination of states that have exactly two pigeons in one box (with the third in a different box), and states in which all three pigeons are in one box as seen in (\ref{eq:InitialState}, \ref{eq:FinalState}). There are in fact no other options, in agreement with the pigeonhole counting principle. We contend that (\ref{eq:Identity1}) provides the quantum analogue of the PCP.

\section{Experiments}
\subsection{Electron Mach-Zehnder Interferometer Experiment}
The authors of \cite{ACPSST} propose an experiment to test the QPC using a Mach-Zehnder interferometer for electrons. In this experiment, three electrons travel along one of two possible paths (see Fig. 1 in \cite{ACPSST}) toward one of two possible detectors. The post-selected state \F is analogous to all three electrons arriving at the same detector. It is presumed that when two or three electrons (pigeons) pass through the same arm (pigeonhole) there will be a repulsion; the strength of interaction is to be tuned so that repulsion is negligible between electrons travelling through different arms. Therefore by observing the position measurements of a large number of runs with the desired post-selected state, one can determine from the spread of position measurements whether any pairs of electrons travelled the same path. The authors predict that no shift in position due to repulsion will be observed (up to first order in an expansion of the repulsion parameter), implying that no two electrons went through the same arm and confirming a violation of the PCP in quantum mechanics.

Aharonov et. al. claim the dynamics of the experiment is governed by the following interaction Hamiltonian (see Eq. (23) in \cite{ACPSST}):
\be
H_{\mbox{\scriptsize int}} = \sum_{a, b= 0}^2 \Pi_{ab} V(r_{ab}),
\ee
where $a < b$ labels the vector space of each electron. This Hamiltonian applies for the period of time $t$ that the electrons travel down the left or right path. In our notation, $|0 \rangle$ and $|1 \rangle$ represent the two arms of the Mach-Zehnder apparatus. The interaction potential is $V(r_{ab})$ where $r_{ab}$ is the distance between electrons.

Assuming the strength of interaction $\epsilon$ is the same between all pairs travelling through the same path we can write this more simply as
\be
H_{\mbox{\scriptsize int}} = \epsilon \Pi .
\ee
{\cb This simplification allows one to go beyond the first order approximations provided in previous articles, and allows us to see the time-dependent, oscillatory interference between terms in the amplitude calculations.}

Time evolution is generated by the exponential of the Hamiltonian. Using (\ref{eq:Identity2}) we have
\be
e^{-i t H_{\mbox{\scriptsize int}} } &=& e^{  -i \epsilon t } e^{ -2 i \epsilon t P }
\ee
since $P$ commutes with the identity operator. Because $P$ is a projector ($P = P^2$) we can find the exact sum of the exponential series involving $P$:
\bea
e^{ -2i \epsilon t P } &=& \mathbb{1} - 2 i \epsilon t P   + \frac{(2i \epsilon t )^2}{2} P + \cdots \none
&=& \mathbb{1}+(e^{-2i \epsilon t}-1)P .
\eea
Now using equation (\ref{eq:Identity1}) to substitute for the identity operator, the time evolution operator can be written as:
\bea
e^{-i t H_{\mbox{\scriptsize int}} } 
&=& e^{  -i \epsilon t }\Pi+(e^{-3i \epsilon t}-3e^{  -i \epsilon t })P \label{eq:Evolution1} 
\label{eq:Evolution2}
\eea

As an interesting aside,  notice (\ref{eq:Evolution2}) shows oscillations in the relative amplitudes of the $\Pi$ and $P$ terms. More to the point, (\ref{eq:Evolution1}) shows that the first order approximation of the time evolution operator is $\mathbb{1} - \epsilon t \Pi$, and that the $P$ operator only contributes at second order and higher. Thus, when $\epsilon t$ is small enough, the Mach-Zehnder experiment provides an experimental measurement of the amplitude
\be
\label{approx}
\left| \langle +i +i +i | e^{-i t H_{\mbox{\scriptsize int}} } | +++ \rangle \right|^2 = 
  \left| \langle +i +i +i | +++ \rangle \right|^2 - 2 \epsilon t{\mathcal Re}\langle +i +i +i | \Pi | +++ \rangle  + \mathcal{O}\left(\epsilon^2 t^2 \right),
\ee
and if there is no deflection observed in position measurements to first order in $\epsilon t$, there must not have been any contribution from the first order term (as predicted by theory) and therefore no two electrons could have travelled the same path. {\cb Note that this proposed experiment is not analogous what was done in \cite{AHM, Chen}.}

\subsection{Predicted Amplitudes}
We now calculate the transition amplitudes generated by the evolution operator (\ref{eq:Evolution1}) between initial state $| + + + \rangle $, and the eight possible final states with each qubit measured in the $|\! \pm \! i \rangle$ basis. Toward this we first find:
\be
\langle (\pm i)_c (+i)_b (+i)_a | \Pi_{ab} | + + + \rangle &=& 0 \label{Pi1} , \\
\langle (\pm i)_c (-i)_b (-i)_a | \Pi_{ab} | + + + \rangle &=& 0 \label{Pi2},\\
\langle (\pm i)_c (+ i)_b (- i)_a | \Pi_{ab} | + + + \rangle &=& \frac{e^{\mp i \pi / 4}}{\sqrt{8}} \label{Pi3}.
\ee
Considering all permutations of qubit labels $a, b, c$, the above equations consider all possible final states. We also find
\be
\langle (\pm i)_c (\pm i)_b (\pm i)_a |P| + + + \rangle &=& \frac{e^{\pm i \pi / 4}}{\sqrt{32}}, \label{P1} \\
\langle (\pm i)_c (\pm i)_b (\mp i)_a | P | + + + \rangle &=& \frac{e^{\mp i \pi / 4}}{\sqrt{32}}. \label{P2}
\ee

The calculations above allow us to find the transition amplitudes for each possible end state. We find
\be
\left| \langle (\pm i)_c (\pm i)_b (\pm i)_a | \exp\left(-i t H_{\mbox{\scriptsize int}} \right) | +++ \rangle \right|^2
&=& \left| \langle (\pm i)_c (\pm i)_b (\pm i)_a | \Pi+(e^{-2i \epsilon t}-3)P  | +++ \rangle \right|^2 \none
&=& \frac{4 - 3 \cos^2 \epsilon t}{8}
\label{amp1}
\ee
which oscillates between $1/8$ and $1/2$ as a function of $\epsilon t$. Considering permutations of $a, b, c$, there are two unique amplitudes found here. Similarly we find that
\be
\left| \langle (\pm i)_c (\pm i)_b (\mp i)_a | \exp\left(-i t H_{\mbox{\scriptsize int}} \right) | +++ \rangle \right|^2
&=& \left| \langle (\pm i)_c (\pm i)_b (\mp i)_a | \Pi+(e^{-2i \epsilon t}-3)P  | +++ \rangle \right|^2 \none
&=& \frac{\cos^2 \epsilon t}{8}
\label{amp2}
\ee
which oscillates between $0$ and $1/8$ as a function of $\epsilon t$. Considering permutations of $a, b, c$, there are six unique amplitudes found here.  All together, one can easily see that the sum of all amplitudes is equal to 1 as it must be, and that to first order, all terms have an equal probability of $1/8$. 

{\cb These are the full amplitudes predicted by the Mach-Zehnder experiment; the interaction strength $\epsilon t$ has an affect only at orders two and higher. Aharonov et. al. choose the first order expansion in order to isolate the term involving only the $\Pi$ operator (see \ref{approx}). The experiments described in \cite{Chen} and \cite{AHM} confirm this term, and separately, the second order term. }

\subsection{Quantum processor experiments}
{\cb Here we describe the information theory behind what is done in \cite{AHM, Chen}, and describe an analogous experiment done a quantum processor.  We measure the transition amplitudes corresponding to certain individual terms in the expansion of (\ref{approx}), namely:
\be
\left| \langle +i +i +i | \Pi_{ab} | +++ \rangle \right|^2 \qquad \mbox{and} \qquad \left| \langle +i +i +i | P | +++ \rangle \right|^2 .
\label{amp3}
\ee
The results of \cite{AHM, Chen} confirm that the first amplitude vanishes and that the second is non-zero, within experimental error.
Here we reproduce these results using the five-qubit quantum ``Yorktown'' processor available through the IBM Quantum Experience cloud service \cite{IBMQ}. We note that previous authors \cite{HDSP} show how a variety of similar experiments could be carried out on quantum processors.}

On an IBM quantum processor, the experimental steps of preparing an initial state, acting with intermediate operators, and performing final measurements can be written as a \textit{quantum circuit}. In these constructs, qubits are labelled with an integer and for our purposes $q[0]$, $q[1]$ and $q[2]$ will play the role of the pigeons. These qubits begin in the state $|000\rangle$ (by coding convention) and are then prepared in the state \I by applying a Hadamard gate to each qubit, i.e. $| \! +++\rangle = H \otimes H \otimes H |000\rangle$.

The `pigeonholes' or `boxes' are the basis states $|0\rangle$ and $|1 \rangle$. Combinations of controlled-not (Cx) gates are connected to ancillary qubits in a way which allows non-destructive measurement of $\Pi_{ab}$ and $P$, as explained below when more details of the algorithm are given. Note that a single Cx gate uses two qubits, one as the control and the other as the target, and is used to entangle or disentangle the states. The target qubit is flipped on the $|1 \rangle$ branch of the control qubit superposition, and left unchanged on the $|0 \rangle$ branch. For example, writing the Cx gate as $\mbox{Cx}(\mbox{control}, \mbox{target})$ and applying this to a general two-qubit state (with $q[0]$ on the right) we obtain
\be
\mbox{Cx}(0, 1)\left( a|0 0 \rangle + b|0 1\rangle + c |1 0\rangle + d |1 1\rangle \right) = a|0 0 \rangle + b|1 1\rangle + c |1 0\rangle + d |1 1\rangle .
\ee
The control qubit is unaffected by this operation.

The measurement operation on an IBM quantum processor is done with respect to the $\left\{|0\rangle, |1\rangle \right\}$ basis by convention. In order to measure the final state of the pigeons in the $|\! \pm \! i \rangle$ basis, one must add rotations of $\pi / 2$ radians about the x-axis on the Bloch sphere before measurement. This is done using $\mbox{Rx}(\pi/2)$ gates which satisfy
\be
|0\rangle = \mbox{Rx}(\pi/2) |\! + \! i\rangle, \qquad \qquad |1\rangle = \mbox{Rx}(\pi/2) |\! - \! i\rangle.
\ee

\subsection{Quantum processor results}
We now describe the quantum circuits, analogous to the experiments done in \cite{AHM, Chen}, which perform intermediate measurements of $\Pi_{01}$ and $P$ between the initial and final states in question. Single qubit gates on the ``Yorktown'' quantum processor are quoted to have an error rate of less than 0.1\% and two qubit gates are quoted to have an error rate of less than 2\% \cite{IBMQ}. More significantly, quantum states of the qubits are not perfectly maintained because they are subject to the processes of energy relaxation and dephasing. We will see that these factors contribute to significant error in the results presented below.

\subsubsection{Measurement of $\Pi_{01}$}

\begin{figure}[htb!]
\begin{center}
\includegraphics[width=0.8\linewidth]{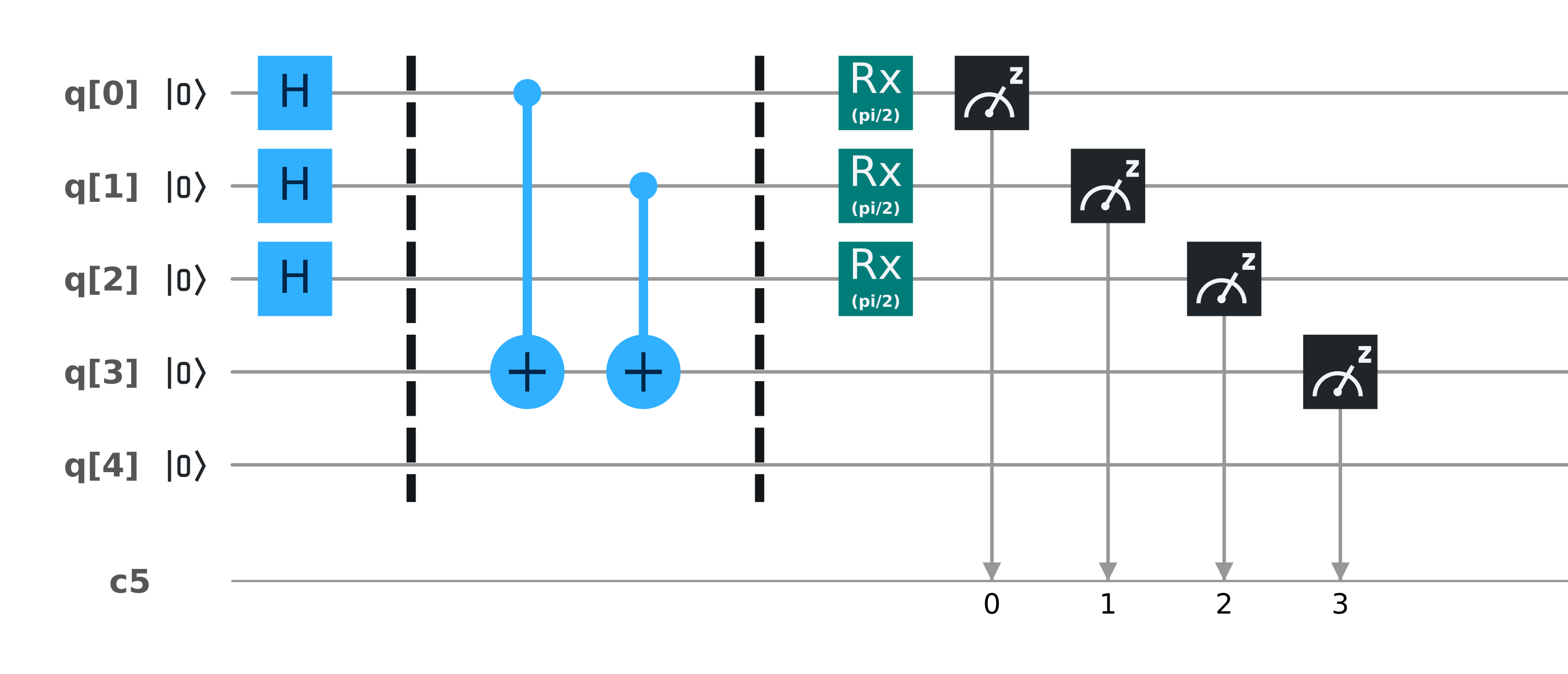}
\caption{A five qubit quantum circuit to check whether qubits $q[0]$ and $q[1]$ are in the same state of the $\left\{|0\rangle, |1\rangle \right\}$ basis. Reading from left to right, the circuit shows five qubits initially in the $|0\rangle$ state with vertical dashed lines (barriers) separating the circuit into three parts: 1) Hadamard gates (denoted by a blue square with an `H') are applied to the `pigeon' qubits $q[0]$, $q[1]$ and $q[2]$ so that these are each in the state $|+\rangle$; 2) a Cx$(0,3)$ and a Cx$(1,3)$ gate are applied with $q[3]$ as the target to detect whether $q[0]$ and $q[1]$ are in the same box. 3) pigeons $q[0]$, $q[1]$ and $q[2]$ are measured in the $| \! \pm \! i\rangle$ basis (by first applying a rotation Rx with angle $\pi/2$ before measurement) and the ancilla $q[3]$ is measured in the $\left\{|0\rangle, |1\rangle \right\}$ basis. The fifth qubit ($q[4]$) is unused in this circuit. The line labelled ``c5'' along the bottom represents the classical five bit register used to hold measurement values. (The actual circuit run on the IBM Yorktown processor was the same as above but with $q[2]$ and $q[3]$ interchanged in order to use the minimum number of gates (and reduce error) considering that controlled-not gates are only possible between certain pairs of qubits.)}
\label{Pi-circuit}
\end{center}
\end{figure}

To measure the amplitudes predicted by (\ref{Pi1} -- \ref{Pi3}) we construct the quantum \textit{circuit} shown in Fig. \ref{Pi-circuit}. The vertical dashed lines are "barriers" which split the circuit into three parts: 1) initial state preparation; 2) application of $\mbox{Cx}$ gates; 3) measurement of pigeons in the $| \!\pm \! i \rangle$ basis and the ancillary qubit in the $\left\{|0\rangle, |1\rangle \right\}$ basis. The Cx gates serve to entangle pigeons $q[0]$ and $q[1]$ with the ancilla $q[3]$ such that $q[3]$ is measured to be $0$ when $q[0]$ and $q[1]$ are in the same box; conversely, $q[3]$ is measured to be $1$ when $q[0]$ and $q[1]$ are in different boxes.

\begin{figure}[htb!]
\begin{center}
\includegraphics[height=2.2in]{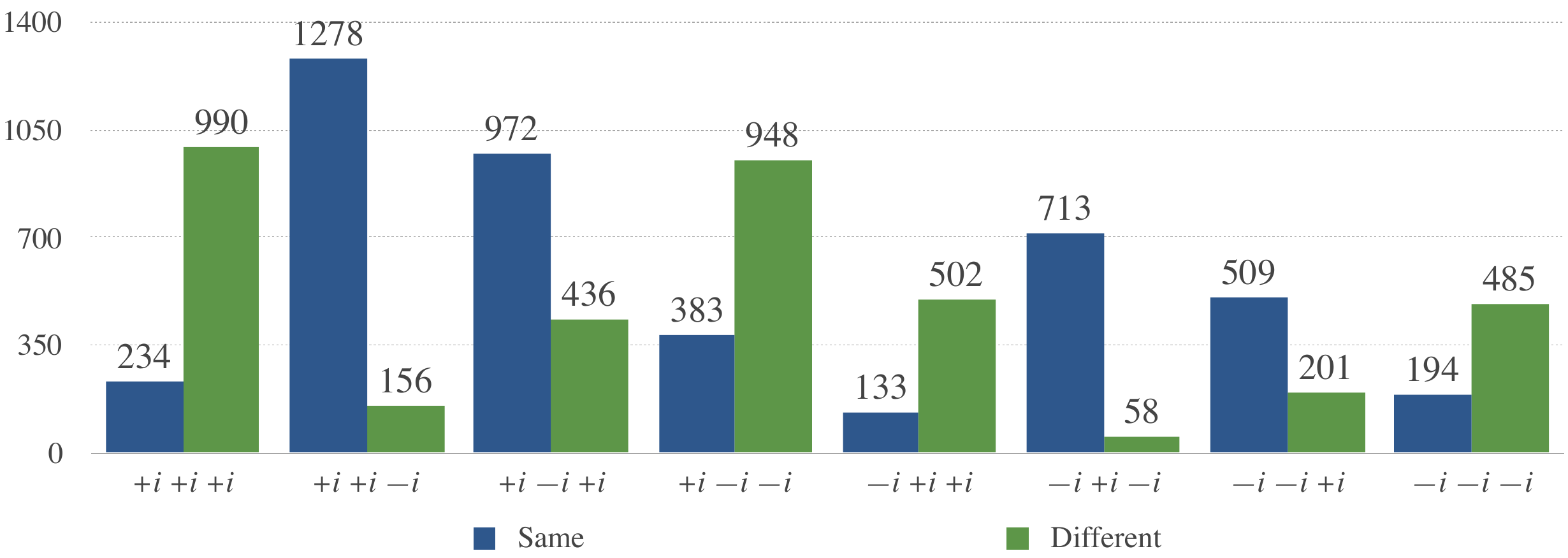}
\caption{Results from 8192 ``shots'' of the quantum circuit shown in Fig. \ref{P-circuit} on the IBM Yorktown processor. For each end state, the histogram charts the counts where $q[3]$ was measured to be zero (pigeons $(q[0],q[1])$ in the same box) and counts where $q[3]$ was measured to be one (pigeons $(q[0],q[1])$ in different boxes). For each post-selected state, the bar with fewer counts is predicted with zero probability, and are only found due to error in the quantum processor.}
\label{Pi-chart}
\end{center}
\end{figure}

This experiment was run a total of 8192 times and the results are presented in Fig. \ref{Pi-chart}. With equal probabilities, the expected count for each post-selected state is 1024 and the associated standard deviation is $\sqrt{1024} = 32$. Notice the total counts for each state are off by a large amount, up to more than 10 standard deviations in some cases. Within each post-selected ensemble, events which are predicted to occur with zero probability are none-the-less found in numbers up to 31\% of the counts for that post-selection. However, events which are predicted to occur do significantly outnumber those which are disallowed by a margin of at least 2.5 times in all cases, agreeing with the amplitudes predicted by (\ref{Pi1} -- \ref{Pi3}).

\subsubsection{Measurement of $P$}

\begin{figure}[htb!]
\begin{center}
\includegraphics[width=0.95\linewidth]{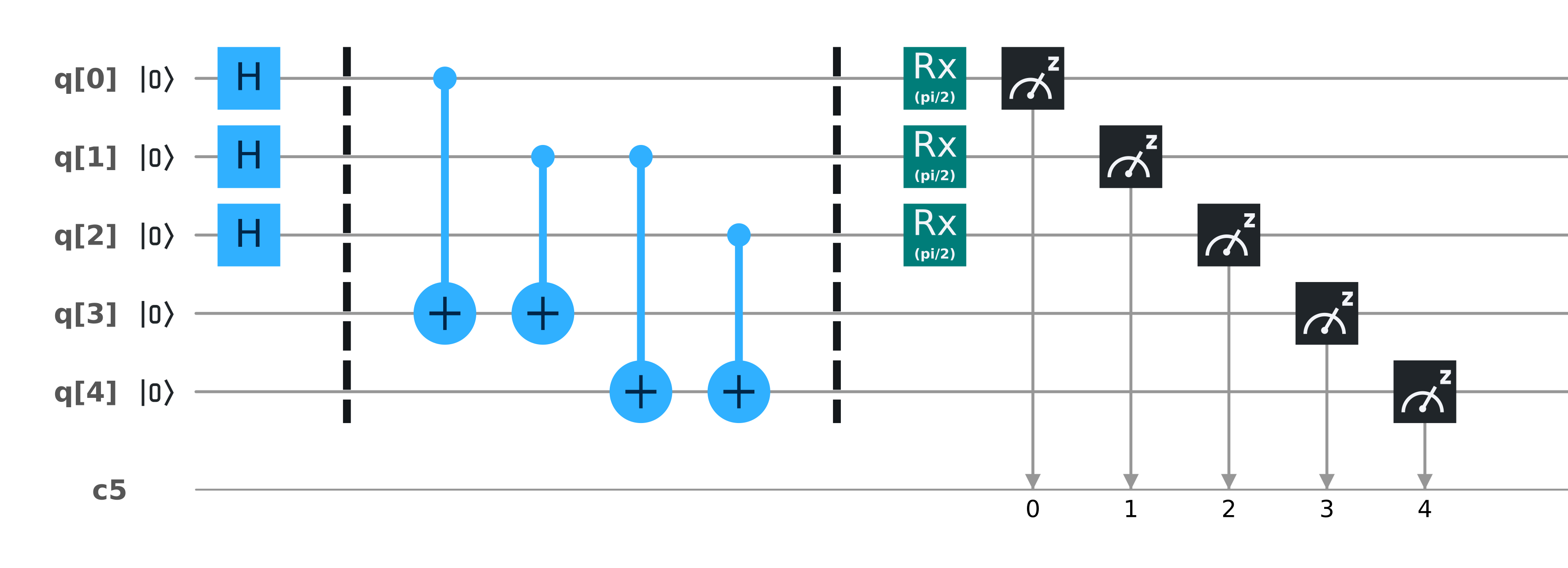}
\caption{A five qubit quantum circuit to check whether the $q[0]$ and $q[1]$ are in the same state of the $\left\{|0\rangle, |1\rangle \right\}$ basis, and also whether the pair $q[1]$ and $q[2]$ are in the same state of the $\left\{|0\rangle, |1\rangle \right\}$ basis. Reading from left to right, the circuit shows five qubits initially in the $|0\rangle$ state with barriers separating the circuit into three parts: 1) Hadamard gates are applied to the "pigeon" qubits $q[0]$, $q[1]$ and $q[2]$ so that these are each in the state $|+\rangle$; 2)  $\mbox{Cx}(0,3)$ and $\mbox{Cx}(1,3)$ gates are applied, followed by $\mbox{Cx}(1,4)$ and $\mbox{Cx}(2,4)$ gates so that measuring ancilla $q[3]$ and $q[4]$ will reveal whether pairs $(q[0], q[1])$  or $(q[1], q[2])$ are in the same box; 3) pigeons $q[0]$, $q[1]$ and $q[2]$ are measured in the $|\pm i\rangle$ basis and the ancilla qubits $q[3]$ and $q[4]$ are measured in the $\left\{|0\rangle, |1\rangle \right\}$ basis. (In order to minimize the number of gates used and the associated error that controlled-not gates are not possible between all qubit pairs, the actual circuit run was the same as above but with relabelled qubit indices: $0 \to 1, 1 \to 2, 2 \to 3, 3 \to 0$.}
\label{P-circuit}
\end{center}
\end{figure}

The circuit shown in Fig. \ref{P-circuit} is designed to measure the amplitudes predicted by (\ref{P1}, \ref{P2}). Here two pairs of $\mbox{Cx}$ gates are used to measure whether pigeons ($q[0]$, $q[1]$) are in the same box (using ancilla $q[3]$), as well as if ($q[1]$, $q[2]$) are in the same box (using ancilla $q[4]$). 

\begin{figure}[htb!]
\begin{center}
\includegraphics[width=0.8\linewidth]{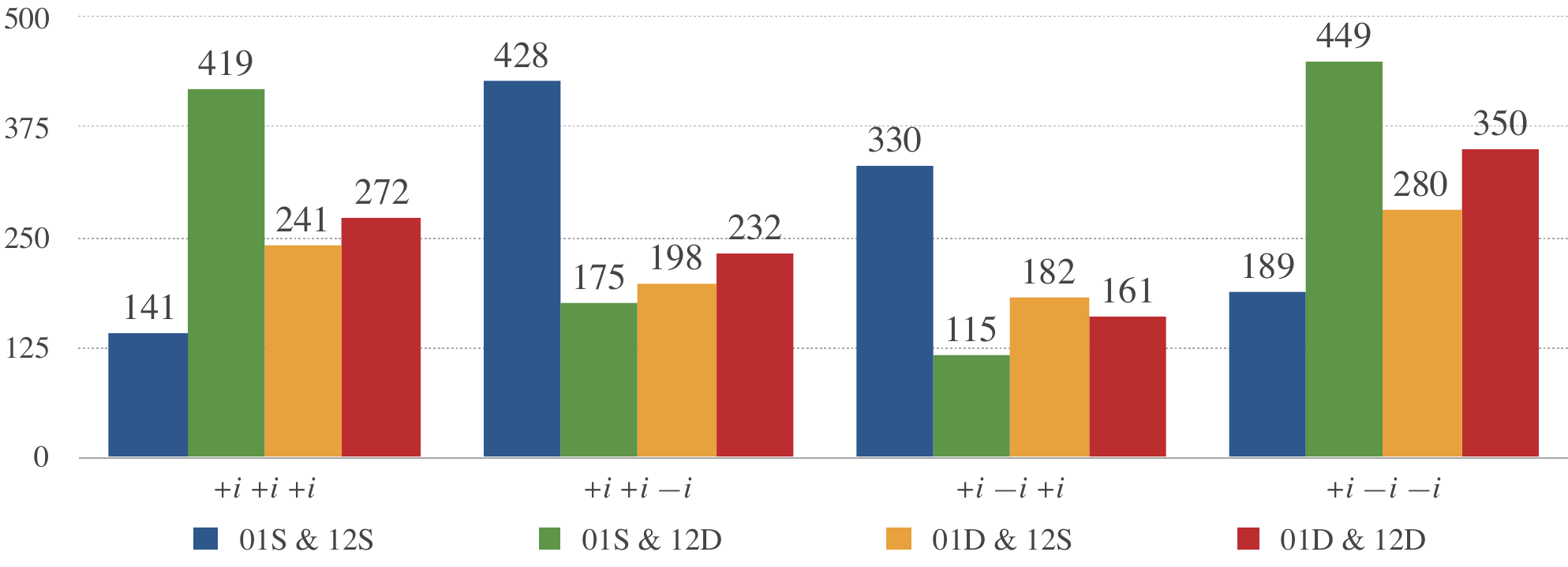}\\[10pt]
\includegraphics[width=0.8\linewidth]{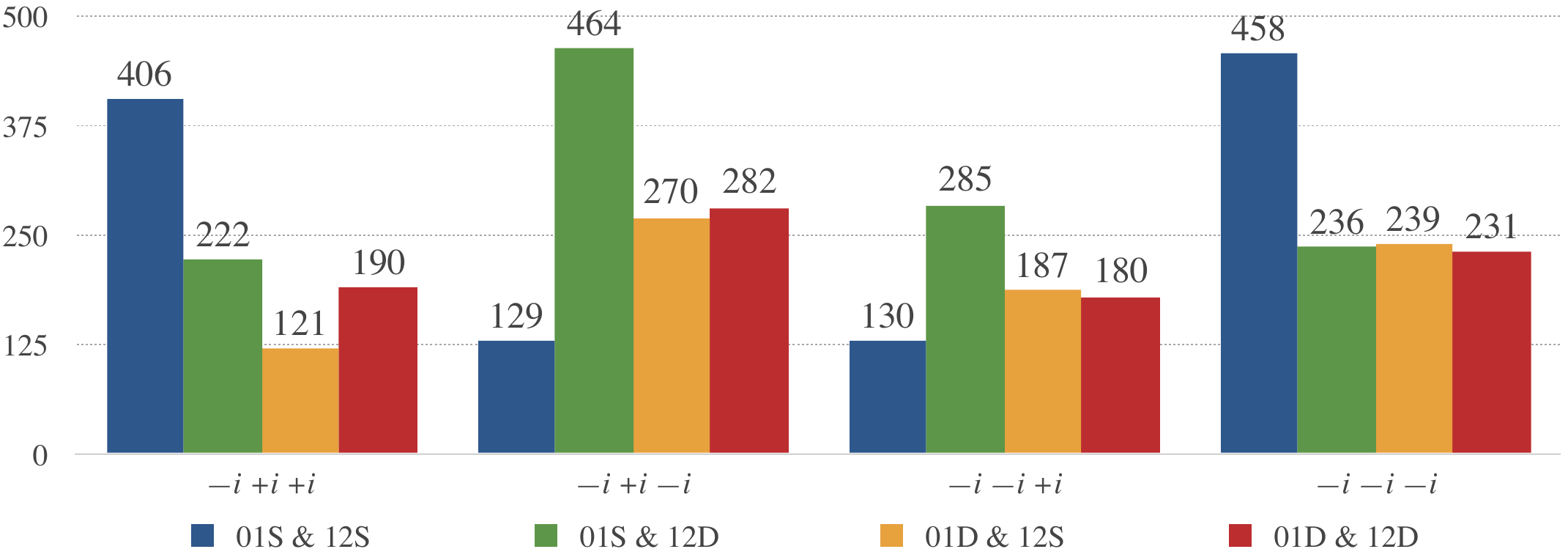}
\caption{Results from 8192 ``shots'' of the quantum circuit shown in Fig. \ref{P-circuit} on the IBM Yorktown processor. For each end state the histogram charts the number of counts where pairs $(q[0], q[1])$ and $(q[1], q[2])$ are found in the same or different boxes, where ``01S'' denotes $(q[0], q[1])$ in the same box, ``01D'' denotes $(q[0], q[1])$ in different boxes, etc.}
\label{P-chart}
\end{center}
\end{figure}

This circuit was also run 8192 times and the results are presented in Fig. \ref{P-chart}. The expected total counts for each post-selected state is again 1024. Also, each of the 32 possible cases (considering possible $P$ measurement outcomes) is predicted with equal probability equal probability for an expectation of 256 counts. Considering the standard deviation of each case, significant deviations are again observed. Within error, these results support the predicted amplitudes in (\ref{P1}, \ref{P2}).

It is enlightening to note that when measuring only $\Pi_{01}$ as in Fig. {\ref{Pi-chart}, it is not possible to find the end state \F (except due to experimental error). However, when applying both $\Pi_{01}$ and then $\Pi_{12}$, it is now possible to find the end state \F. This exemplifies the strange nature of quantum mechanics, and how the entire five-qubit state is playing a role --- it is not possible to distinguish the three pigeon qubits from the ancilla used for measurement. In the quantum computing realm, it is known that such temporary ancilla must be ``uncomputed'' (disentangled) before they can be used again, in order to not affect the other qubits in an entangled quantum state.

\section{Discussion}
Let us now return to the basis of the QPC as described in \cite{ACPSST}. It is argued that (\ref{eq:PiTransition}) implies
the only cases in which the final measurement can find the state \F are those in which the intermediate measurement of any two particles $a$ and $b$ finds that they are in different boxes \footnote{Note that if one thinks of the quantum wavefunction as describing the statistics for an ensemble of similarly prepared experiments, this statement assigns values to observables related to the positions of pigeons in individual members of that ensemble.}.
Their point is that measuring (i.e. projecting onto) the final state \F cancels all branches of the initial state in which $a$ and $b$ are in the same box.
Due to the symmetry of the initial and final states, this is true for any choice of $\{a, b\} \in \{0, 1, 2\}$. Aharonov et. al. conclude that ``given the above pre- and post-selection, we have three particles in two boxes, yet no two particles can be found in the same box...". However, this conclusion does follow from the arguments.

First, note that the operator identity (\ref{eq:Identity1}) makes clear that the Hilbert space is spanned by states that have either exactly two or exactly three pigeons in the same hole. The vanishing of $\langle +\! i +\! i +\! i | \Pi_{ab} |\! +++ \rangle$ is due to interference between the two branches of the superposition where pigeons $a$ and $b$ are in the same hole. But this is not the only choice of operator to project onto branches where the pigeons are in the same hole, and the other choices lead to non-vanishing amplitudes. For example $\langle +\! i +\! i +\! i | P | +++ \rangle$ with all three pigeons in the same box does not vanish, nor do  projections of two pigeons into a \textit{particular} box, $|0\rangle$ or $|1\rangle$. This point was already raised in the arXiv version of the letter by Svensson \cite{Svensson}. We now present new arguments in the context of hidden variable theories to show that the QPC is simply a novel example of quantum weirdness that is by now well known and understood.

The three projection operators $\Pi_{ab}$ all mutually commute so that simultaneous eigenstates exists. It is permissible therefore to assign specific values (they must be eigenvalues) of these observables to each element of the ensemble that projects to the final state. But as in the case of Bell's inequalities and Bell's more general proof of the non-viability of local hidden variables theories \cite{Bell1966, Mermin1993}, there are restrictions to the values $v(\Pi_{ab})$ that these measurements can take. The identity operator (\ref{eq:Identity2}) shows that the value of $\Pi$ can never be zero \footnote{In fact, the eigenvalues of $\Pi$ are 1 and 3 so any measurement would necessarily find one of these values.} since for any realizable state, if two of $v(\Pi_{ab})$ are zero then the third must have unit eigenvalue (and $v(P)$ must be zero). Therefore this identity forbids the assignment of zero to all three $v(\Pi_{ab})$ simultaneously, which shows the fallacy in the statement ``no two pigeons can be found in the same box''. If one measures $\Pi_{01}$, say, then  this changes the statistics of possible outcomes for $\Pi_{02}$ and $\Pi_{12}$ by eliminating from the ensemble all elements for which both $v(\Pi_{02})$ and $v(\Pi_{12})$ are zero. One can verify that if pigeons 1 and 2 are known to be in different boxes and pigeons 1 and 3 are known to be in different boxes, then pigeons 2 and 3 are necessarily in the same box. Thus, despite the vanishing of the transition amplitude (\ref{eq:PiTransition}), there is no way to create a physical state in which the observed, or unobserved, values of three of the commuting projection operators $\Pi_{ab}$ vanish, i.e. in which no two pigeons are in the same box.}

The initial state vector is a linear combination of all basis vectors, each having at least two pigeons in the same box, and
each with an equal probability of being measured. Post-selection of the state \F is equivalent to projection onto this state; if a $\Pi$ operation is applied in the intermediate region, then no elements of the ensemble reach the final state \F, which is what the experiments confirm. To conclude that the PCP is violated, Aharonov et. al. assert knowledge of each $v(\Pi_{ab})$ without actually applying these operators. Projection onto the final state, without the intermediate $\Pi$ operations leaves a linear combination of all basis vectors differing in phase but still with equal measurement probabilities. What then can we learn about the relative positions of the pigeons before the final projection from the fact that measuring $\Pi_{ab}$ in an intermediate state \textit{would have} eliminated the states with $a$ and $b$ in the same box? Nothing, because the projection onto the final state does not commute with $\Pi_{ab}$, so that the eigenvalues of the two projection operators cannot be determined simultaneously.

The important lesson from all this therefore is not that no two pigeons are in the same box, but that quantum mechanics does not allow one to assign values to all three relevant observables in the absence of specific measurements in this scenario. The situation is analogous to 
what happens when Bell's inequalities are violated (see \cite{Mermin1981} for a beautiful exposition).
In this case, one starts with an ensemble of pairs of emitted particles, $A$ and $B$, in a state of zero total spin. The spin correlation between the particles allows one to predict the spin of particle $B$, say, along any given axis after the spin of $A$ is measured along the same axis, despite the absence of causal interactions between the two particles after emission.  However, Bell's inequalities imply that one cannot, in general, assign simultaneous values to spin measurements along three different axes. The proof in this case is again by contradiction: if one assumes that spins of both particles in a given element of the ensemble have  definite values along three suitably chosen axes immediately after the particles are first emitted, irrespective of any measurements, then these values can only match the statistics predicted by quantum mechanics if some of the spin configurations occur a negative number of times in the ensemble. That is, the probability of some configurations must be negative. This is a violation of basic counting principles similar to the QPC.
 
Note that in the present case, the inability to assign the desired specific values to the operators $\Pi_{ab}$ is not due to the fact that the states on which they are acting are entangled. Instead, the projection operators $\Pi_{ab}$ themselves entangle the particles and obey identities that forbid these assignments.  The QPC  provides a novel and elegant pedagogical framework for understanding the constraints imposed on reality by quantum mechanics but does not, however, allow one to conclude that there are no two pigeons in the same box in violation of the PCP, just as Bell's inequalities do not allow one to claim that the sum of two positive numbers is negative.\\[5pt]

{\cb
\section{Conclusion}

The authors of \cite{ACPSST} calculate the transition amplitude
\be
 \left| \langle + i +\! i +\! i | \Pi | +++ \rangle \right|^2 = 0.
 \label{eq:TAzero}
\ee 
This implies three qubits beginning in state \I and ending in state \F are not in the same $z$-basis state \textit{when $\Pi$ is applied intermediately}. They then make a leap of logic to conclude that no two qubits are in the same $z$-basis state \textit{even when $\Pi$ is not applied intermediately}. {This conclusion is not supported by theory since the operator $\Pi$ does not commute with projection onto the state \F,  and so one cannot create a quantum state that simultaneously has definite values of both.} The fact that experiments \cite{AHM, Chen} verify the above equation (\ref{eq:TAzero}) has no bearing on the validity of the conclusion.

{More important than clarifying the flaw in the argument of \cite{ACPSST}, we have provided a proof that the PCP is not violated in quantum mechanics.} The operators $\left\{ \mathbb{1}, \Pi, P \right\}$ form a commuting set and satisfy the equation
\be
\mathbb{1} = \Pi -2P.
\ee
Since they can be simultaneously measured, their values must also satisfy this relationship for any quantum state. The authors of \cite{ACPSST} implicitly imply that $v(\Pi) = v(P) = 0$ in stating that no two or three pigeons are in the same hole, but this would violate the identity
\be
1 = v(\Pi) -2 v(P),
\ee
which must be obeyed by measurements of any of the possible quantum states. Therefore the PCP is not violated in quantum mechanics, the experiments do not demonstrate a previously unknown structure of quantum mechanics, and our understanding of separability and correlations is not in question.
}

\vspace{11pt}
{\bf Acknowledgments}\\[5pt]
G.K. is grateful to the Natural Sciences and Engineering Research Council for funding.


\end{document}